# Continuous-time Quantum Walks on a Cycle Graph

Dmitry Solenov and Leonid Fedichkin

Center for Quantum Device Technology,
Department of Physics, Clarkson University, Potsdam, NY 13699-5721
Electronic addresses: solenov@clarkson.edu, leonid@clarkson.edu

We present analytical treatment of quantum walks on a cycle graph. The investigation is based on a realistic physical model of the graph in which decoherence is induced by continuous monitoring of each graph vertex with nearby quantum point contact. We derive the analytical expression of the probability distribution along the cycle. Upper bound estimate to mixing time is shown.

Quantum walks have been widely discussed recently as a promising technique for development of quantum algorithms [1,2]. Both, discrete-time quantum coined walks and continuous-time quantum walks have been argued to give an algorithmic speed-up with respect to its classical counterparts [3]. Unlike common discrete-time quantum algorithms [4] that are very sensitive to environmental quantum noise [5], quantum walks show some promise in dealing with decoherence processes. Numerical studies of discrete-time quantum walks on a cycle and hyper-cube have shown that small amount of decoherence may be useful [6]. In this Letter we present theoretical investigation of continuous-time quantum walks on a uniform cycle graph, $C_N$. We derive the expression for the probability distribution and obtain the upper-bound estimate to mixing time.

In our investigation, the cycle is represented by a ring-shaped array of identical tunnel-coupled quantum dots (QDs), see Fig. 1. The walks are performed by an electron initially placed in one of the dots. Each dot is continuously monitored by an individual point contact (PC), which introduces decoherence to electron's evolution as discussed in Ref. 7. The analytical expression for the probability distribution is obtained for a cycle of arbitrary size, i.e. the number of nodes may be large. The latter property allows studying dynamics and mixing on the large graphs avoiding usual limitations on size arising in numerical simulations [6].

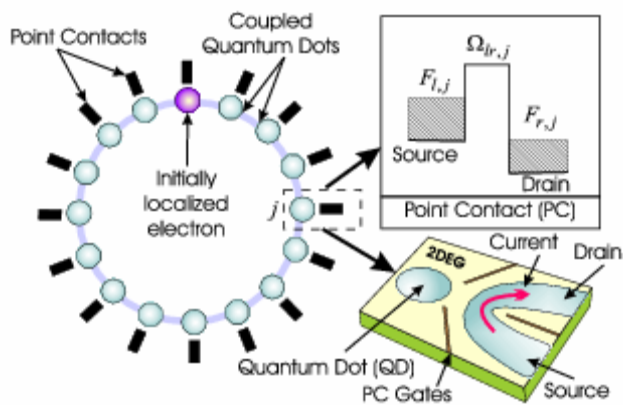

**Figure 1.** Continuous-time quantum walks architecture: ring of quantum dots, each of which is monitored by the corresponding point contact that introduces decoherence. $F_{l,j}$ and $F_{r,j}$ are chemical potentials of source and drain of $j$-th point contact. Presence of an electron in $j$-th quantum dot affects source-to-drain tunneling amplitude $\Omega_{lr,j} \to \Omega_{lr,j} + \delta\Omega_{lr,j}$ of $j$-th point contact.

The QD cycle with "attached" PCs can be, in principle, fabricated with the help of gate-engineering technique in semiconductor heterostructures [8]. It allows forming QDs and PCs electrostatically by placing metal dates on the structure with two-dimensional electron gas (2DEG). By changing potential on the gates one can allocate areas of 2DEG creating necessary confinement profile. The simplest example of such a structure, containing two QDs was investigated experimentally in Ref. 9. Our key assumptions are as follows: identical PCs are formed far enough from QD-structure so that the tunneling between them is negligible; Coulomb interaction between electrons in QD and PC is taken into account.

We begin with formulating the basic equations of our model. The Hamiltonian of an electron placed in the QD-cycle is

$$H_{cycle} = \frac{1}{4}\sum_{j=0}^{N-1}\left(c_{j+1}^{\dagger}c_j + c_j^{\dagger}c_{j+1}\right), \qquad (1)$$

where $c_j^{\dagger}$ ($c_j$) are creation (annihilation) operators for an electron on site $j$; $N$ is the number of QDs in the cycle, and $c_N \equiv c_0$. We renormalize the time for convenience, so that it becomes dimensionless, and all the amplitudes further on are given in terms of hopping amplitude between neighboring QDs.

The point contact, placed next to each QD, consists of two reservoirs of electrons: source and drain that are coupled through the potential barrier shaped by PC gates, see Fig. 1. The Hamiltonian of $j$-th PC can be written as

$$H_{PC,j} = \sum_l E_{l,j} a_{l,j}^+ a_{l,j} + \sum_r E_{r,j} a_{r,j}^+ a_{r,j} + \sum_{lr} \Omega_{lr,j}\left(a_{l,j}^+ a_{r,j} + a_{r,j}^+ a_{l,j}\right), \qquad (2)$$

where $a_{l,j}^{\dagger}(a_{l,j})$ and $a_{r,j}^{\dagger}(a_{r,j})$ are creation (annihilation) operators in the left (source) and right (drain) reservoirs of $j$-th PC. $\Omega_{lr,j}$ are the tunneling amplitudes between states $l$ and $r$ of $j$-th PC. In our discussion we consider all electrons to be spin-less fermions. Source and drain reservoirs are kept at zero temperature with chemical potentials $F_{l,j}$ and $F_{r,j}$. By allowing weak Coulomb interaction between electrons in PC and QD we observe the presence of the electron in $j$-th QD as it changes the tunneling amplitude through the barrier of adjoined PC, i.e. effectively $\Omega_{lr,j} \rightarrow \Omega_{lr,j} + \delta\Omega_{lr,j}$, so that $\delta\Omega_{lr,j}$ represents the rise of the potential barrier in PC when the corresponding QD is occupied. The correction is assumed to be the small comparing to the other amplitudes in the problem. This process introduces weak measurement on the electron in each node of the graph, and, therefore, results in some loss of coherence in electron evolution. Summarizing the above discussion, we produce the following correction to PC Hamiltonian (2),

$$H_{int,j} = \sum_{lr}\delta\Omega_{lr,j} c_j^+ c_j \left(a_{l,j}^+ a_{r,j} + a_{r,j}^+ a_{l,j}\right). \qquad (3)$$

The total Hamiltonian is

$$H = H_{cycle} + \sum_{j=0}^{N-1}\left(H_{PC,j} + H_{int,j}\right). \qquad (4)$$

In our investigation we assume that all PCs are identical, and that hopping amplitudes $\Omega_{lr,j}$ are only weakly dependent on states $l,r$, which allows to replace $\Omega_{lr,j}$ and $\delta\Omega_{lr,j}$, as well as $F_{l,j}(F_{r,j})$, by their averages: $\bar{\Omega}$, $\delta\bar{\Omega}$ and $\bar{F}_l(\bar{F}_r)$. Considering continuous measurement of an electron in double-well potential by point contact described above, S.A. Gurvitz have shown [7] that for the case of large bias voltages $\bar{F}_l - \bar{F}_r$, the evolution of the reduced density matrix traced over all states of source and drain electrons is given by Bloch-type rate equations. Applied to our model this technique yields the following equation for the reduce density matrix

$$\frac{d}{dt}\rho_{\alpha\beta} = \frac{i}{4}\left(\rho_{\alpha\beta+1} - \rho_{\alpha+1\beta} - \rho_{\alpha-1\beta} + \rho_{\alpha\beta-1}\right) - \Gamma\left(1-\delta_{\alpha\beta}\right)\rho_{\alpha\beta}, \qquad (5)$$

where $\alpha,\beta$ number the sites on the cycle, running from 0 to $N-1$; $\Gamma = \delta\bar{\Omega}^2\left(\bar{F}_r - \bar{F}_l\right)^2 f_S f_D$; and $f_S(f_D)$ stand for density of states in source (drain) reservoirs. We also set $\hbar = 1$ for convenience.

For further discussion, it is convenient to introduce real variables, defining

$$\rho_{\alpha\beta} \equiv i^{\alpha-\beta} S_{\alpha\beta}. \qquad (6)$$

Considering (6) we obtain

$$\frac{dS_{\alpha\beta}}{dt} = \sum_{\mu,\nu=0}^{N-1} \left( L_{\alpha\beta}^{\mu\nu} + U_{\alpha\beta}^{\mu\nu} \right) S_{\mu\nu} , \qquad (7)$$

where $\alpha, \beta, \mu, \nu$ run from $0$ to $N-1$, and we have $L_{\alpha\beta}^{\mu\nu}$ and $U_{\alpha\beta}^{\mu\nu}$ defined as

$$L_{\alpha\beta}^{\mu\nu} = \frac{1}{4} \left( \delta_{\alpha,\mu}\delta_{\beta,\nu-1} + \delta_{\alpha,\mu-1}\delta_{\beta,\nu} - \delta_{\alpha,\mu}\delta_{\beta,\nu+1} - \delta_{\alpha,\mu+1}\delta_{\beta,\nu} \right) , \qquad (8)$$

$$U_{\alpha\beta}^{\mu\nu} = -\Gamma \delta_{\alpha,\mu}\delta_{\beta,\nu}\left(1 - \delta_{\alpha,\beta}\right) . \qquad (9)$$

As mentioned earlier, we initialize the system by localizing the electron in one of the quantum dots and allow it to evolve, spreading all over the cycle. Therefore, the reduced density matrix elements at $t = 0$ are set as follows

$$\rho_{\alpha\beta}(0) = S_{\alpha\beta}(0) = \delta_{\alpha,0}\delta_{\beta,0} . \qquad (10)$$

Condition (10) simply states that the electron is initially localized in dot "0". The choice of initial condition in form (10) is convenient for further calculations and, in fact, is quite general. Indeed, the symmetry of the system with respect to cyclic rotations allows to construct the solution to the reduced density matrix for any classical, i.e. with zero off-diagonal elements, initial distribution. The solution for the desired initial distribution is given by the superposition, as

$$\sum_{j=0}^{N-1} C_j \rho_{\alpha+j\beta+j}(t) , \qquad (11)$$

where $C_j$ represent initial probability distribution along the cycle.

Equations (7) can be solved perturbatively in low decoherence (quantum) regime, considering $\Gamma N \ll 1$. Zero-order solution is given as an expansion on the eigenvectors of $L_{\alpha\beta}^{\mu\nu}$, defined by

$$\sum_{\mu,\nu=0}^{N-1} L_{\alpha\beta}^{\mu\nu} V_{\mu\nu}^{(mn)} = \lambda_{(mn)}^{0} V_{\alpha\beta}^{(mn)} , \qquad (12)$$

where $0 \leq m, n \leq N - 1$. From equation (12), after some algebra, one can show that eigenvalues $\lambda_{(mn)}^0$ are

$$\lambda_{(mn)}^0 = i \sin\frac{\pi(m+n)}{N} \cos\frac{\pi(m-n)}{N} , \qquad (13)$$

and eigenvectors $V^{(mn)}$ are given by

$$V_{\mu\nu}^{(mn)} = \frac{1}{N} e^{\frac{2\pi i}{N}(m\mu+n\nu)} . \qquad (14)$$

Calculations of the corrections require careful investigation of the unperturbed spectrum (13). The analysis of (13) and (14) allows highlighting several important subsets of certain degeneracy which lead to non-zero off-diagonal matrix elements of (9) on the basis of (14). First of all, one can notice the symmetry of (13) with respect to indexes swap, while eigenvectors (14) are clearly affected by such the operation. Hence, for $n \neq m$ we deal with at least two-fold degenerate eigenvalues. Another subset, reveal itself when we consider eigenvalues (13) with $m = n = 0$ or $m + n = N$. Eigenvalues (13) with these relations for indexes are all zeros and yet the corresponding eigenvectors are not the same.

First-order corrections to eigenvalues $\lambda_{(nn)}^0$ of spectrum (13) are given, as one can show, by the diagonal matrix elements of (9) calculated on eigenvectors (14). They equal to $-\Gamma(N-1)/N$. The perturbation removes degeneracy of the first subset introducing $-\Gamma(N-1\pm 1)/N$ to each pair of $\lambda_{(mn)}^0$ with $n \neq m$ and $n+m \neq N$. Corrections to zero eigenvalues are irrelevant to our calculations due to the fact that corresponding eigenvectors are anyway excluded from the final expression by initial condition (10). To show that one can simply analyze the expansion of the right hand-side of (10) in terms of eigenvectors (14) that yields

$$S_{\alpha\beta}(0) = \frac{\delta_{\alpha,\beta}}{N} + \frac{1}{N^2} \sum_{m,n=0}^{N-1} \left(1 - \delta_{m+n,0} - \delta_{m+n,N}\right) \exp\left[\frac{2\pi i (m\alpha + n\beta)}{N}\right] \quad (15)$$

The solution to (7) is naturally formed as

$$S_{\alpha\beta}(t) = \sum_{m,n=0}^{N-1} C_{(mn)} e^{\lambda_{(mn)} t} Y_{\alpha\beta}^{(mn)}, \quad (16)$$

where $Y_{\alpha\beta}^{(mn)}$ are some linear combinations of eigenvectors (14) and $\lambda_{(mn)}$ represent the corrected spectrum. Expansion coefficients $C_{(mn)}$ are completely defined by the form of expression (15). Finally, the solution to (7) is

$$S_{\alpha\beta}(t) = \frac{\delta_{\alpha,\beta}}{N} + \sum_{m,n=0}^{N-1} \frac{1 - \delta_{m+n,0} - \delta_{m+n,N}}{N} \left[\delta_{mn} e^{t\lambda_{(mn)}^0 - \Gamma \frac{N-1}{N} t} + (1-\delta_{mn}) e^{t\lambda_{(mn)}^0 - \Gamma \frac{N-2}{N} t}\right] V_{\alpha\beta}^{(mn)}. \quad (17)$$

The probability distribution, which is given by the diagonal elements of the reduced density matrix (17), considering (6), is

$$P_j(t) = \frac{1}{N} + \sum_{m,n=0}^{N-1} \frac{1 - \delta_{m+n,0} - \delta_{m+n,N}}{N^2} \left[\delta_{mn} e^{-\Gamma \frac{N-1}{N} t} + (1-\delta_{mn}) e^{-\Gamma \frac{N-2}{N} t}\right] \times$$
$$\times \exp\left[it \sin\frac{\pi(m+n)}{N} \cos\frac{\pi(m-n)}{N} + \frac{2\pi i}{N}(m+n)j\right]. \quad (18)$$

Expression (18) is already the result of our investigation. It gives the probability for the electron, initially placed at node "0", to be found on node $j$ at time $t$. The probability distribution is shown in Fig. 2. As one can see, the pattern of coherent walks, Fig. 2A, seems to be almost unaffected when the system is exposed to weak measurement (decoherence), Fig. 2B. The latter, however, suppresses the coherent oscillation pattern introducing effective averaging that leads to onset of uniform distribution.

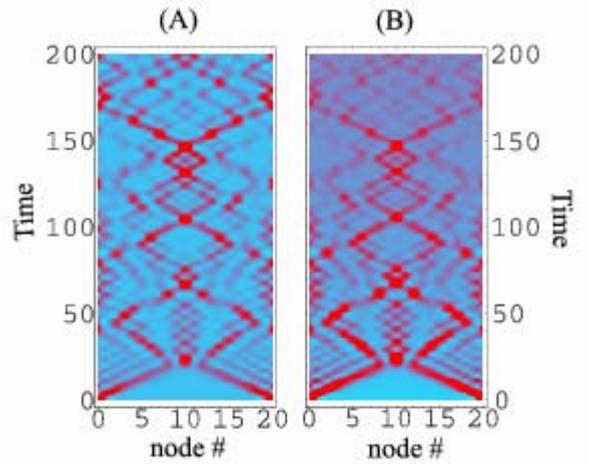

**Figure 2.** Probability distribution along the cycle as function of time and node number, for $N=20$ and $\Gamma=0$ (A), $\Gamma=0.01$ (B). Here $j \in [0, N-1]$ stands for the node number; darker regions denote higher probabilities. Electron is initially placed at $j=0$. The probability distribution of the walks with some decoherence added, (B), converges to uniform, i.e. to $1/N$.

In quantum walks studies, it is often important to analyze the time it takes for the electron, as a walking particle, to spread along the cycle. This is called "mixing time" [2,6,10], and for continuous-time quantum walks is used to describe two types of processes. The first one, called "instantaneous mixing," refers to uniform (or nearly uniform) spread of probability of the walking particle that can happen at some particular moment and is destroyed a moment later [10]. The other, "average mixing," is the decay of time-averaged deviation of the probability distribution from the uniform [6]. In the latter case, time averaging is required to settle down the coherent oscillations of probability which, otherwise, would not converge to any static distribution. In our case the averaging arises naturally from the fact that the electron walking on the cycle is continuously monitored by the environment, i.e. PCs.

Let us briefly discuss how possibly fast mixing on a circle can be. One of the apparent necessary (but not at all sufficient) conditions is for the walking particle to have some nonzero amplitude on each node. Therefore, the wave of probability of the particle localized initially in one of the nodes has to travel all over the cycle at least once. We should note, that for our range of parameters this has already happened by times of order $1/\Gamma$.

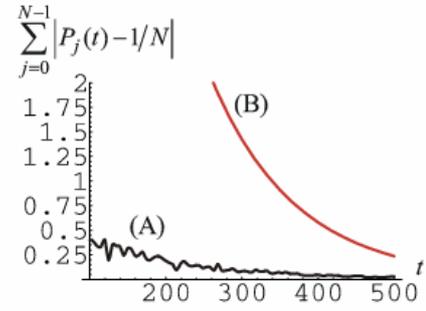

**Figure 3.** The sum of the absolute deviations of probability from uniform, curve (A), with majorizing curve, (B), are shown. Analysis of the latter allows analytical expression for the upper bound estimate to mixing time, which is found to be linear in $N$. Parameters used are as follows: $N = 20$ and $\Gamma = 0.01$.

Further on, we obtain an upper bound estimate based on the solution (18). The mixing time, $t_{mix}$, is defined [6] as the minimum time that satisfies mixing condition

$$\sum_{j=0}^{N-1} \left| P_j(t_m) - \frac{1}{N} \right| \leq \varepsilon , \qquad (19)$$

where $\varepsilon$ is some small dimensionless constant that presets the desired degree of mixing, and $1/N$ stands for the uniform distribution. To find the upper bound estimate, let us analyze the left part of inequality (19). After some algebra with expression (18) one can obtain

$$\left| P_j(t_m) - \frac{1}{N} \right| = e^{-\Gamma \frac{N-2}{N} t_m} \left| S^2(j, t_m/2) - \frac{1}{N} + \frac{e^{-\Gamma t_m/N} - 1}{N} \left[ S(2j, t_m) - \frac{2 - N \bmod 2}{N} \right] \right| , \qquad (20)$$

where

$$S(j,t) = \frac{1}{N} \sum_{n=0}^{N-1} e^{it \sin \frac{2\pi n}{N} + i \frac{2\pi n}{N} j} . \qquad (21)$$

The absolute value of sum (21) is always smaller than or equal to unity, which allows us to majorize (20) as follows

$$\left| P_j(t_m) - \frac{1}{N} \right| \leq e^{-\Gamma \frac{N-2}{N} t_m} \left[ 1 + \frac{1}{N} + \frac{e^{-\Gamma t_m/N} - 1}{N} \left( 1 + \frac{2}{N} \right) \right] . \qquad (22)$$

In Fig. 3 we plot sum of the absolute deviation of probability distribution from the uniform one, curve (A), along with the majorizing expression, curve (B). Substituting (22) into (19) and assuming $N > 2$ we yield the relation

$$N + \frac{2}{N} \leq \varepsilon \left[ \exp(\Gamma t_m/N) \right]^{N-2} + \frac{N+2}{N} \exp(-\Gamma t_m/N) , \qquad (23)$$

which always has a solution at some large $t_m$. The upper bound for the mixing time can be defined as $t_{mix} \leq \min t_m$. The latter minimum is estimated assuming that the last term in (23) is small and using the relation $n^{1/(n-k)} \leq k + 1$ which holds for integers $n > k \geq 2$. As a result we obtain

$$t_{mix} \leq \frac{N}{\Gamma} \ln\left(\frac{4}{\varepsilon}\right) , \qquad (24)$$

which is consistent with the above assumption. As we see, in low decoherence mode the mixing time may not exceed the quantity linear in $N$ for a given $\Gamma$. On the other hand for a fixed size of the cycle the mixing time is expected to decrease as $1/\Gamma$. Eventually, as $\Gamma$ increases one goes to the regime of strong measurement with emerging Zeno effect, where the electron is localized by the measurement itself, which obviously destroys mixing. Observing (24), one may speculate that there must be some optimal value for decoherence parameter $\Gamma$ which corresponds to minimum mixing time for a given size of a cycle. This behavior requires careful investigation and goes beyond the scope of the present paper. We should also note, that instantaneous mixing (if exists) can actually happen much earlier as compared to (24). The mixing time in the latter case is determined, primarily, by the pattern of coherent oscillations.

In conclusion, we have studied quantum walks on a cycle graph, represented by a ring-shape array of quantum dots continuously monitored by individual point contacts, which introduce decoherence. Analytical expression for the probability distribution along the cycle has been obtained for small amount of decoherence. We have shown that at fixed low decoherence rates the upper bound estimate for mixing time has linear dependence on the size of the cycle, while fixing the size, one observes inverse linear dependence on the decoherence rate.

We are grateful to Christino Tamon for helpful discussions. This research was supported by the National Science Foundation, Grant DMR-0121146.